\documentclass[12pt,preprint,preprint,nofootinbib]{revtex4}
 \pdfoutput=1
\usepackage{epsf, array, color}
\usepackage{graphicx}
\usepackage{subfigure}
\usepackage{bbold}
\usepackage{amsmath}
\usepackage{mathtools}
\usepackage{dcolumn}
\usepackage{url}
\usepackage{verbatim}
\usepackage{wrapfig}
\usepackage{amsfonts}
\usepackage{comment}
\usepackage{epigraph}
\usepackage{slashed}
\usepackage[utf8]{inputenc}
\usepackage[linktocpage,colorlinks,urlcolor=blue]{hyperref}
\usepackage{bbm}
\usepackage{cancel}
\usepackage{epsf, array, color}

\newcommand{\nc}{\newcommand}

\newcommand{\bea}{\begin{eqnarray}}
\newcommand{\eea}{\end{eqnarray}}
\newcommand{\nn}{\nonumber \\}

\def\beq{\begin{equation}}
\def\eeq{\end{equation}}
\newcommand{\vp}{\phi}

\def\nn{\nonumber}
\def\gev{\rm GeV}

\newcommand{\C}{\mathcal{C}}
\def\Op{\mathcal{O}}

\nc{\vpj }{\mbox{${\vp^\dag i\,\raisebox{2mm}{\boldmath ${}^\leftrightarrow$}\hspace{-4mm} D_\mu\,\vp}$}}
\nc{\vpjt}{\mbox{${\vp^\dag i\,\raisebox{2mm}{\boldmath ${}^\leftrightarrow$}\hspace{-4mm} D_\mu^{\,I}\,\vp}$}}

\begin{document}
\begin{flushright}

 {\small 
IFT-UAM/CSIC-18-117, PITT-PACC-1816
}\\

\end{flushright}

\title{SMEFT and the Drell-Yan Process at High Energy}

\author{ S.~Dawson$^{(a)}$, P.~P.~Giardino$^{(b)}$, and A.~Ismail$^{(c)}$ }
\affiliation{
\vspace*{.5cm}
  \mbox{$^{(a)}$Department of Physics,\\
  Brookhaven National Laboratory, Upton, N.Y., 11973,  U.S.A.}\\
  \mbox{$^{(b)}$Instituto de F\'isica Te\'orica UAM/CSIC, Universidad Aut\'onoma de Madrid, 28049, Madrid, Spain}\\
  \mbox{$^{(c)}$PITT PACC, University of Pittsburgh, Pittsburgh, PA., 15260, U.S.A.
}\\
 \vspace*{1cm}}

\date{\today}

\begin{abstract}
The Drell-Yan process is a copious source of lepton pairs at high energy and is measured with great precision at the Large Hadron Collider (LHC).  
Barring any new light particles, beyond the Standard Model effects can be studied in  Drell-Yan production 
using an effective field theory.  At tree level, new $4$-fermion interactions dominate, while at one loop operators 
modifying $3$-gauge boson couplings contribute effects that are enhanced at high energy.  We study the sensitivity of the neutral
Drell-Yan process to these dimension-$6$ operators  and  compare the sensitivity to that of $W^+W^-$ pair production at the LHC.

\end{abstract}
\maketitle

\section{Introduction}
\label{sec:intro}
The exploration of the electroweak sector is a major task for the LHC.  Without new low scale particles, the
only tools available for studying deviations from the SM predictions are effective field theories (EFT).  In this approach,
low scale physics is assumed to be sensitive to the presence of higher dimension  operators.  When a complete basis of
these operators is constructed, they will affect predictions for many observables, including in Higgs physics~\cite{Giudice:2007fh,Brivio:2017vri}, gauge
boson pair production~\cite{Baglio:2017bfe,Azatov:2016sqh,Zhang:2016zsp,Grojean:2018dqj,Alves:2018nof}, top quark production~\cite{AguilarSaavedra:2018nen,Degrande:2012gr,Farina:2018lqo}, and many other processes.  The measurements from different processes provide
 complementary information about the parameters of the EFT and potential insights into the underlying UV physics.

In this work, we consider the effects of a consistent EFT analysis on neutral Drell-Yan production.
The Drell-Yan  process is extremely precisely measured at numerous energies, while the Standard Model theoretical predictions exist at NNLO
QCD~\cite{Anastasiou:2003yy,Gehrmann-DeRidder:2016jns,Ridder:2016nkl,Karlberg:2014qua,Hoeche:2014aia}, along with the resummation of the logarithms~\cite{Bizon:2018foh,Alioli:2015toa}.
The QCD corrections have
been combined with NLO electroweak effects~\cite{Dittmaier:2009cr,Baur:2001ze} and implemented in the
FEWZ code~\cite{Li:2012wna,Gavin:2010az,CarloniCalame:2007cd}.  
We study  neutral Drell-Yan production  in the context of  the SMEFT~\cite{Grzadkowski:2010es}, where the Higgs boson is assumed
to be part of an $SU(2)$ doublet.  The effects of the dimension-$6$ SMEFT operators can potentially be of the same magnitude as 
the higher order Standard Model corrections and both need to be considered in precision studies.   New $4$-fermion operators 
can contribute to $q {\overline{q}} \rightarrow l^+l^-$ production at tree level and have been extensively studied in the 
literature~\cite{Falkowski:2017pss,Berthier:2015oma,deBlas:2013qqa,Cirigliano:2012ab,Carpentier:2010ue}. Precision measurements at the $Z$-pole place bounds on the strengths of the non-Standard
Model $4-$fermion operators, and even more stringent constraints come from other low energy measurements including atomic parity violation, deep inelastic scattering and flavor observables.

At the LHC, new information can be gained by looking at high $p_T$ (or $m_{ll}$) events where the
new physics effects are potentially enhanced by contributions of ${\cal {O}}({p_T^2\over \Lambda^2})$.
In  an EFT approach, the NLO corrections to the EFT contributions are also necessary, and new operators that do not contribute at tree level can have measurable effects.  
The QCD corrections to Drell-Yan production in the SMEFT are known~\cite{Alioli:2018ljm}.
The  program of electroweak corrections to the SMEFT is in its infancy, however, with results for 
$H\rightarrow VV$~\cite{Dawson:2018pyl,Dawson:2018liq,Hartmann:2015aia,Dedes:2018seb},
  $H\rightarrow b {\overline b}$~\cite{Gauld:2016kuu,Gauld:2015lmb} and  
$Z\rightarrow f {\overline f}$~\cite{Hartmann:2016pil,Dawson:2018jlg} known.  Here we begin the program of one-loop EFT contributions to Drell-Yan production.  We consider the one-loop contributions
from anomalous $3$-gauge boson interactions, and compare with the sensitivity to these interactions in $W^+W^-$ pair production. The sensitivity of Drell-Yan production
to oblique corrections at high energy has  also been studied in Ref.~\cite{Farina:2016rws}. We find that while Drell-Yan provides additional information, the impact of anomalous $3$-gauge boson interactions is generally more easily observable in $W^+W^-$ pair production. Additionally, 4-fermion operators which were not considered in Ref.~\cite{Farina:2016rws} affect the Drell-Yan process at tree level, and unless they are set to zero, as in a universal theory, they can overwhelm the impact of loop corrections from other operators.

In Section~\ref{sec:basics}, we review the SMEFT and write the leading order amplitude for Drell-Yan production. Section~\ref{sec:nlo} shows the results of our NLO calculation involving SMEFT operators. Then, in Section~\ref{sec:results} we demonstrate the impact of SMEFT operators on kinematic distributions in Drell-Yan production, and estimate the reach of the LHC in probing these operators. Section~\ref{sec:concl} contains our conclusions.

\section{Basics}
\label{sec:basics}

In the SMEFT,  new physics is described
by a tower of operators,
\begin{equation}
{\cal L}={\cal L}_{SM}+\Sigma_{k=5}^{\infty}\Sigma_{i=1}^n {\mathcal{C}_i^k\over \Lambda^{k-4}} O_i^k\, .
\label{eq:lsmeft}
\end{equation}
The dimension-$k$ operators are constructed from SM fields and the
new beyond the SM  (BSM)  physics effects reside in the coefficient functions, $\C_i^k$.  
For large $\Lambda$,  it is sufficient to retain only the lowest dimensional operators.
The operators have been classified in several different bases, which 
are related by the equations of motion \cite{Buchmuller:1985jz,Grzadkowski:2010es,Hagiwara:1993ck,Giudice:2007fh}.  
In this paper we will use the Warsaw basis of 
Ref. \cite{Grzadkowski:2010es} and the convenient implementation of Ref. \cite{Dedes:2017zog}.

Only a few operators contribute to 
the neutral Drell-Yan process, $f {\overline f}\rightarrow e_p^+e_p^-$,  at tree level ($p$ is a generation index).  There are new $4$-fermion operators, along with operators that shift the tree level relationships among the 
parameters \cite{Falkowski:2015fla,Brivio:2017bnu}.  The operators relevant for Drell-Yan production  at tree level, 
along with the operator $\Op_W$  that is the focus of the next section, are given in Table \ref{tab:opdef}. We define $q$ and $l$ to be
the $SU(2)_L$ quark and lepton doublets, respectively.

\begin{table}[t] 
\centering
\renewcommand{\arraystretch}{1.5}
\begin{tabular}{||c|c||c|c||c|c||} 
\hline \hline
$\Op_W$                & 
$\epsilon^{IJK} W_\mu^{I\nu} W_\nu^{J\rho} W_\rho^{K\mu}$ &    
$\Op_{\vp D}$   & 
$\left(\vp^\dag D^\mu\vp\right)^* \left(\vp^\dag D_\mu\vp\right)$ &
$\Op_{\vp WB}$ &
$\left(\vp^\dag \tau^I\vp\right)^*W_{\mu\nu}^IB^{\mu\nu} $
\\
\hline
$\underset{p,r}{\Op_{\vp l}^{(3)}}$      & 
$(\vpjt)(\bar l'_p \tau^I \gamma^\mu l'_r)$&
$\underset{p,r,s,t}{\Op_{l q}^{(1)}}$      & 
$(\bar l'_p \gamma_\mu l'_r)(\bar q'_s \gamma^\mu  q'_t)$  &
$\underset{p,r,s,t}{\Op_{l q}^{(3)}}$  &
$(\bar l'_p \gamma_\mu \tau^I l'_r)(\bar q'_s \gamma^\mu \tau^I q'_t)$  
\\
\hline
$\underset{p,r,s,t}{\Op_{ q e}}$      & 
$(\bar q'_p \gamma_\mu q'_r)(\bar e'_s \gamma^\mu  e'_t)$  &
$\underset{p,r,s,t}{\Op_{e u}}$            & 
$(\bar e'_p \gamma_\mu e'_r)(\bar u'_s \gamma^\mu  u'_t)$  &
 $\underset{p,r,s,t}{\Op_{e d}}$            & 
$(\bar e'_p \gamma_\mu e'_r)(\bar d'_s \gamma^\mu  d'_t)$ 
 \\
 \hline
$\underset{p,r,s,t}{\Op_{l u}}$            & 
 $(\bar l'_p \gamma_\mu l'_r)(\bar u'_s \gamma^\mu  u'_t)$  &
 $\underset{p,r}{\Op_{l d}}$            & 
$(\bar l'_p \gamma_\mu l'_r)(\bar d'_s \gamma^\mu  d'_t)$  &
    $\underset{p,r,s,t}{\Op_{ll}}$        & 
    $(\bar l'_p \gamma_\mu l'_r)(\bar l'_s \gamma^\mu l'_t)$  
    \\
\hline \hline
\end{tabular}
\caption{Dimension-6 operators relevant for our study
  (from \cite{Grzadkowski:2010es}). For brevity we suppress fermion
  chiral indices $L,R$. $I=1,2,3$ is an $SU(2)$ index.  $p,r,s,t=1,2,3$ are generation indices. \label{tab:opdef}}
\end{table}

The operators in Table \ref{tab:opdef}  change the form of the kinetic terms  in the  gauge sector,
\bea
\mathcal{L}&=&-\frac14 W^{I,\mu\nu} W_{\mu\nu}^I-\frac14 B^{\mu\nu}B_{\mu\nu}\nn\\
&+&\frac1{\Lambda^2}\bigg(\C_{\phi W} (\phi ^\dagger \phi )W^{I,\mu\nu} W_{\mu\nu}^I+\C_{\phi B} (\phi ^\dagger \phi )B^{\mu\nu} B_{\mu\nu}+\C_{\phi WB} (\phi ^\dagger \tau^I \phi )W^{I,\mu\nu} B_{\mu\nu} \bigg).
\eea
We  define ``barred'' fields, ${\overline W}_\mu\equiv (1-\C_{\phi W} v^2/\Lambda^2)W_\mu$ and ${\overline B}_\mu\equiv (1-\C_{\phi B}v^2/\Lambda^2)B_\mu$ and ``barred'' gauge couplings,  ${\overline g}_2\equiv (1+\C_{\phi W} v^2/\Lambda^2)g_2$ and ${\overline g}_1\equiv (1+\C_{\phi B}v^2/\Lambda^2)g_1$ so that ${\overline W}_\mu {\overline g}_2= W_\mu g_2$ and ${\overline B}_\mu {\overline g}_1= B_\mu g_1$. The ``barred'' fields defined in this way have their kinetic terms properly normalized and preserve the form of the covariant derivative.  
The masses of the W and Z fields are then, \cite{Dedes:2017zog,Alonso:2013hga}:
\bea
M_W^2&=&\frac{{\overline g}_2^2 v^2}4,\nn\\
M_Z^2&=&\frac{({\overline g}_1^2+{\overline g}_2^2) v^2}4+\frac{v^4}{\Lambda^2}\left(\frac18 ({\overline g}_1^2+{\overline g}_2^2) \C_{\phi D}+\frac12 {\overline g}_1{\overline g}_2\C_{\phi WB} \right).
\eea

Dimension-6 operators  contribute to the decay of the $\mu$ lepton at tree level, changing the relation between the 
vev, $v$, and the Fermi constant $G_\mu$ obtained from the measurement of the $\mu$ lifetime,
\begin{eqnarray}
G_\mu& =&\frac1{\sqrt{2} v^2}-\frac1{2\sqrt{2}\Lambda^2}
(\underset{1,2,2,1}{\C_{ll}}+\underset{2,1,1,2}{\C_{ll}})+{\sqrt{2}\over 2\Lambda^2}
(\underset{1,1}{\C_{\phi l}}^{(3)}+\underset{2,2}{\C_{\phi l}}^{(3)})
\nonumber \\
&\equiv &\frac1{\sqrt{2} v^2}-\frac1{\sqrt{2}\Lambda^2}\C_{ll}+{\sqrt{2}\over \Lambda^2}\C_{\phi l}^{(3)},
\end{eqnarray}
where we assume flavor universality of the coefficients in the last line above. 

 We choose the $G_\mu$ scheme, where we take the physical  input parameters to be,
\begin{eqnarray}
G_\mu&=&1.1663787(6)\times 10^{-5} \gev^{-2}\nonumber \\
M_Z&=&91.1876\pm .0021\gev\nonumber \\
M_W&=&80.385\pm .015~\gev
\end{eqnarray}
   
Using an input basis of $M_W,~M_Z$ and $G_\mu$\cite{Brivio:2017bnu}, the effective fermion-$Z/\gamma$ interactions are,
\begin{eqnarray}
L&=&{2M_W\over v}\sqrt{1-{M_W^2\over M_Z^2}}Q_f
\biggl\{1-{c_W\over s_W}v^2 \C_{\vp WB} -{1\over 4} {c_W^2\over s_W^2}v^2\C_{\vp D}\biggr\}{\overline{f}}\gamma^\mu f A_\mu\nonumber\\
&&+{2M_Z\over v}\biggl\{T_3^f-Q_f(1-{M_W^2\over M_Z^2})(1-{c_W\over s_W}v^2\C_{\vp WB})
\nonumber \\ &&
-{v^2\over 4}\biggl(T_3^f-Q_f(1+{M_W^2\over M_Z^2})\biggr)\C_{\vp D}
-{v^2\over 2} (\C_{\vp f}^{(1)}-2T_3^f\C_{\vp f}^{(3)})
\biggr\}
{\overline{f}}\gamma^\mu P_L f Z_\mu\nonumber \\
&&+{2M_Z\over v}\biggl\{-Q_f(1-{M_W^2\over M_Z^2})(1-{c_W\over s_W}v^2\C_{\vp WB})
\nonumber \\
&&+{v^2\over 4}Q_f(1+{M_W^2\over M_Z^2})\C_{\vp D}-{v^2\over 2}\C_{\vp f}
\biggr\}
{\overline{f}}\gamma^\mu P_R f Z_\mu\, ,
\end{eqnarray}
where $c_W=M_W/M_Z$, $s_W=\sqrt{1-c_W^2}$, $T_3^f=\pm {1\over 2}$, $P_{L,R}={(1\pm\gamma_5)\over 2}$, and $f=q,l$ for left-handed
fermions and $f=u,d,e$ for right-handed fermions.  (We omit the generation indices for simplicity).

The tree level SM result for $ q_i {\overline q}_i\rightarrow  e_p^+e_p^-$ receives corrections from s-channel
$Z/\gamma$ exchange,
\begin{eqnarray}
A_{XY}^{\rm SM}=M_{XY}\biggl\{ {4M_W^2\over v^2}s_W^2
{Q_lQ_q \over s}+{g_{qX} g_{lY}\over s-M_Z^2}\biggr\}\, ,
\end{eqnarray}
where $X,Y=L,R$ ,
\begin{eqnarray}
g_{fL}&=&{2M_Z\over v} \biggl[T_3^f-Q_f\biggr(1-{M_W^2\over M_Z^2}\biggr)\biggr]\nonumber \\
g_{fR}&=& {2M_Z\over v}\biggl[-Q_f\biggr(1-{M_W^2\over M_Z^2}\biggr)\biggr]\, ,
\end{eqnarray}
and,
\begin{eqnarray}
M_{XY}&\equiv& \biggl({\overline {f}}_r\gamma_\mu P_X f_r\biggr)\biggl( {\overline {e}}_p\gamma^\mu P_Y e_p\biggr) \, .
\end{eqnarray}

The tree level SMEFT amplitudes instead read,

\begin{eqnarray} 
A_{XY}^{\rm SMEFT}=A_{XY}^{\rm SM} - \frac{v^2}{\Lambda^2}M_{XY}\biggl\{ {8M_W^2\over v^2}(s_W c_W C_{\vp WB}+\frac{c_W^2}4 C_{\vp D})
{Q_lQ_q \over s}+\frac12 C_{\vp D}{g_{qX} g_{lY}\over s-M_Z^2} \nonumber \\
-\frac{2 M_Z}{v}(s_W c_W C_{\vp WB}+\frac{c_W^2}2 C_{\vp D}){g_{qX}Q_l+ g_{lY}Q_q\over s-M_Z^2}
\biggr\}+A_{XY,q}^{4-fermions}\, ,
\end{eqnarray}
where
\begin{eqnarray}
A_{LL,q}^{4-fermions}&=&\frac{M_{XY}}{\Lambda^2}(\underset{2,2,1,1}{C_{lq}^{(1)}}-2T_3^q \underset{2,2,1,1}{\C_{lq}^{(3)}})
\nonumber \\  A_{LR,q}^{4-fermions}&=&\frac{M_{XY}}{\Lambda^2}(\underset{1,1,2,2}{\C_{qe}})\nonumber \\
A_{RL,u(d)}^{4-fermions}&=&\frac{M_{XY}}{\Lambda^2}(\underset{2,2,1,1}{\C_{lu(d)}})
\nonumber \\ A_{RR,u(d)}^{4-fermions}&=&\frac{M_{XY}}{\Lambda^2}(\underset{2,2,1,1}{\C_{eu(d)}})\, .
\end{eqnarray}
The $4$-fermion operators give contributions that grow with energy relative to the SM contributions and the phenomenological effects
have been examined in Ref. \cite{Berthier:2015oma,deBlas:2013qqa}.  Our results are in agreement with these references. 

\section{NLO Amplitudes}
\label{sec:nlo}

\begin{figure}
\centering
\includegraphics[width=0.8\textwidth]{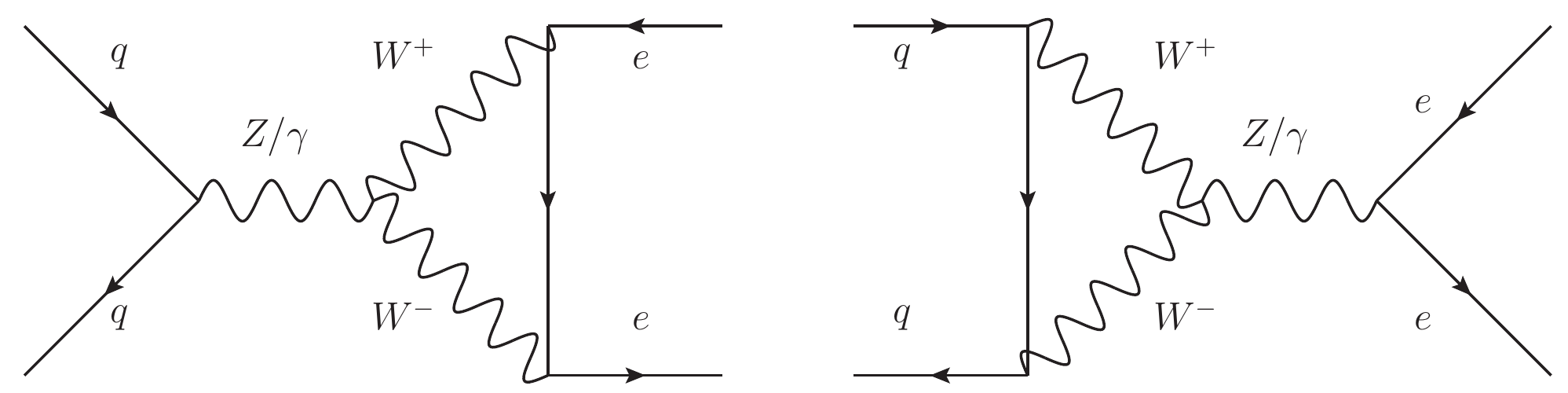}
\caption{Vertex corrections proportional to $\C_W$.}
\label{fig:diagrams}
\end{figure}

At one loop, there are contributions to Drell-Yan from new operators not contributing at tree level.  We focus on $O_W$.  
This operator is particularly interesting because it is strongly restricted  from $f {\overline f} \rightarrow W^+W^-$ both
at LEP and at the LHC as its effects grow with energy.  The sensitivity to $\C_W$ from global fits to LHC measurements and to LEP data has been
found in Ref.~\cite{Alves:2018nof,Biekotter:2018rhp,Almeida:2018cld,Bjorn:2016zlr}, and is roughly,
\begin{equation}
-0.17 <\C_W\biggl({1~\mathrm{TeV}\over\Lambda^2}\biggr)^2<0.18 \, .
\label{eq:cwlimit}
\end{equation}
It has been speculated~\cite{Farina:2018lqo} that because of the large cross section and precision of the measurements that high energy
Drell-Yan could also yield a precise determination of $\C_W$. The fits of Refs. \cite{Alves:2018nof,Biekotter:2018rhp,Almeida:2018cld,Bjorn:2016zlr} 
 include the measurement of the Drell-Yan process on the $Z$-peak.

The diagrams of Fig.~\ref{fig:diagrams} give contributions to the left-hand amplitudes.
The complete one-loop amplitudes proportional to $\C_W$ are available in the \href{https://quark.phy.bnl.gov/Digital_Data_Archive/dawson/drellyan_18}{online archive} associated with our results and the energy enhanced (relative to the SM)
contributions are,
 \begin{eqnarray} 
 A_{LL,u}^{NLO}&=&A_{LL,u}^{SMEFT}\biggl(1 -\biggl[{3 sv^2\over \Lambda^2 M_Z^2 (1+2c_W^2)}\biggr]\biggl\{
{g^3 \C_W\over 32 \pi^2}\biggr\}\biggr)\nonumber \\
A_{LL,d}^{NLO}&=&A_{LL,d}^{SMEFT}\biggl(1+ \biggl[{3 sv^2\over \Lambda^2 M_Z^2 (1-4c_W^2)}\biggr]\biggl\{
{g^3  \C_W\over 32 \pi^2}\biggr\}\biggr)\, .
\end{eqnarray}
These amplitudes include the NLO shifts of the input parameters~\cite{Chen:2013kfa,Dawson:2018jlg} as well as the one-loop diagrams proportional to $\C_W$, consisting of both vertex and propagator corrections. When the contributions from the input parameter shifts and the one-loop diagrams are added together, the divergences cancel completely in the NLO amplitudes as expected.

\section{Results}
\label{sec:results}

We now use the amplitudes of Sections~\ref{sec:basics} and \ref{sec:nlo} to calculate the effects of SMEFT operators on kinematic distributions in Drell-Yan production.
Current and future precision measurements of $p p \to e_p^+ e_p^-$ will constrain not only $4$-fermion operators involving quarks and leptons, but also purely bosonic operators that contribute at loop level. We concentrate on the effect of the latter, that would dominate in a 
universal theory~\cite{Wells:2015cre} or one in which the sizes of the coefficients of BSM $4$-fermion operators involving 1st and 2nd generation quarks and leptons are small.

\begin{figure}
\centering
\includegraphics[width=0.49\textwidth]{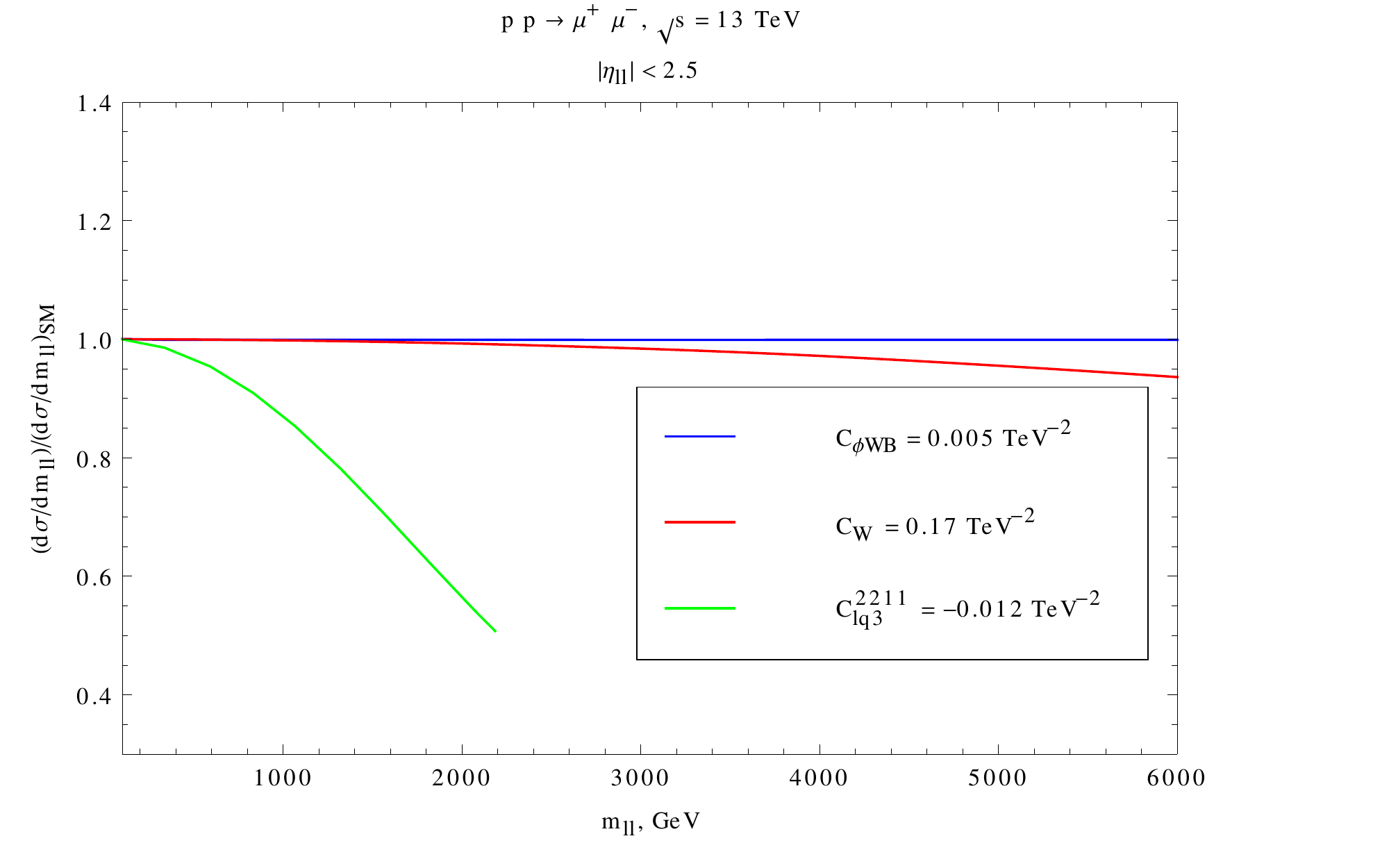}
\includegraphics[width=0.49\textwidth]{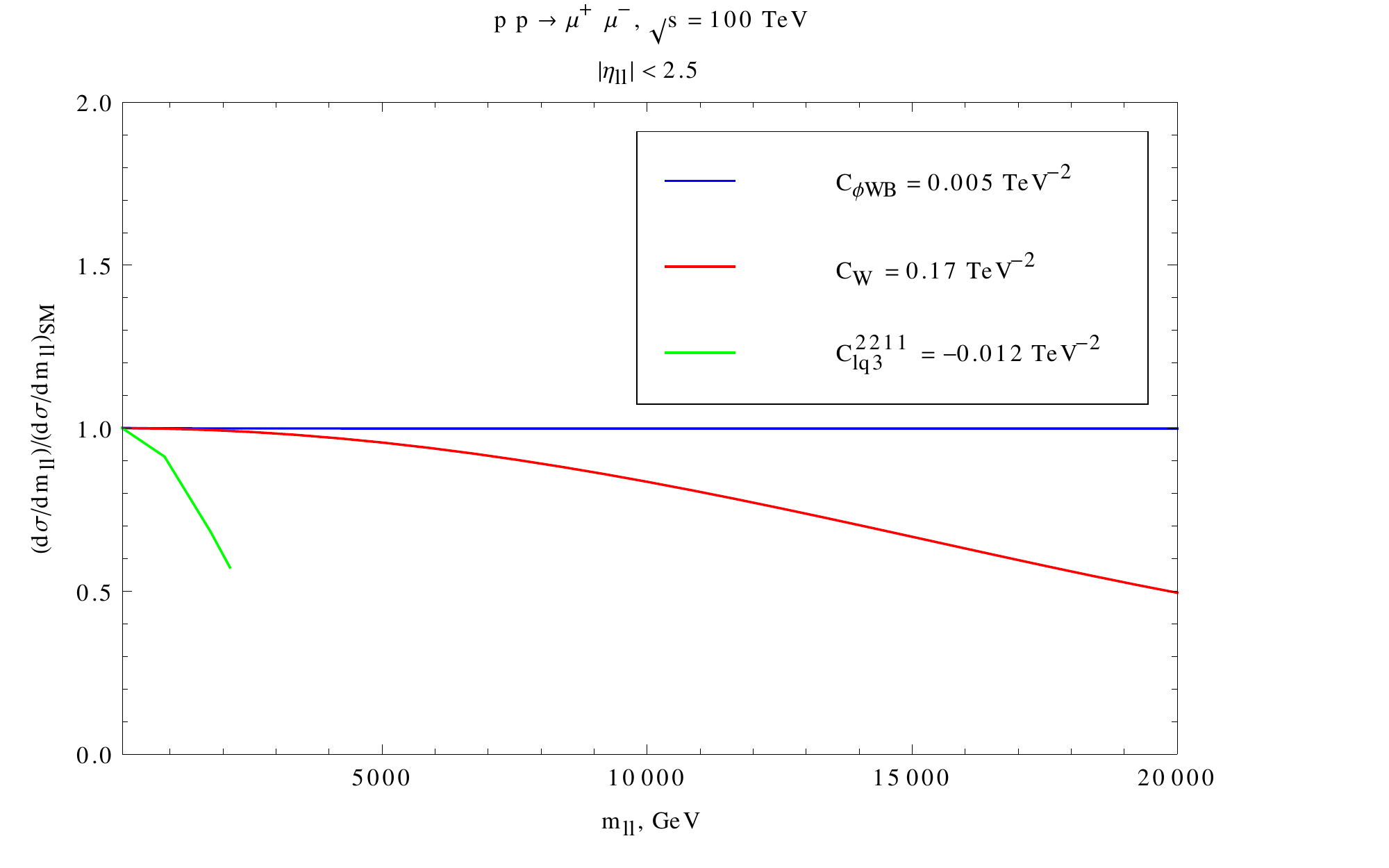}
\caption{The ratio of the differential cross section as a function of dilepton mass in the SMEFT to that in the SM. In each curve, one operator coefficient is turned on at a time, while the others are set to zero. The sizes of the operators are taken to be at their current bounds. The left (right) plot shows the distribution for 14 (100) TeV. The results for the 4-fermion operator are cut off where the BSM contribution becomes similar in magnitude to the SM amplitude, as detailed in the text.}
\label{fig:dsdmll}
\end{figure}

\begin{figure}
\centering
\includegraphics[width=0.49\textwidth]{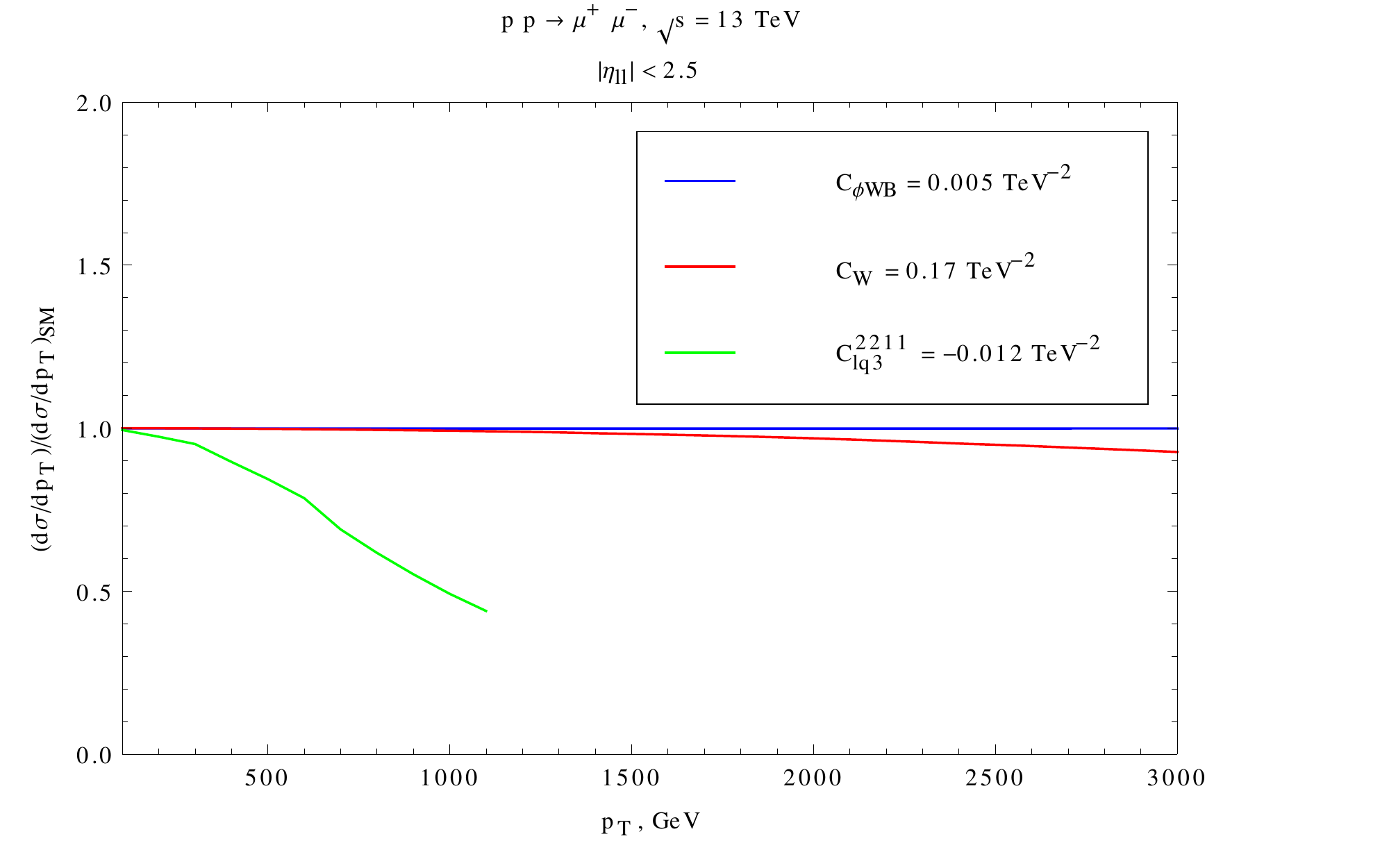}
\includegraphics[width=0.49\textwidth]{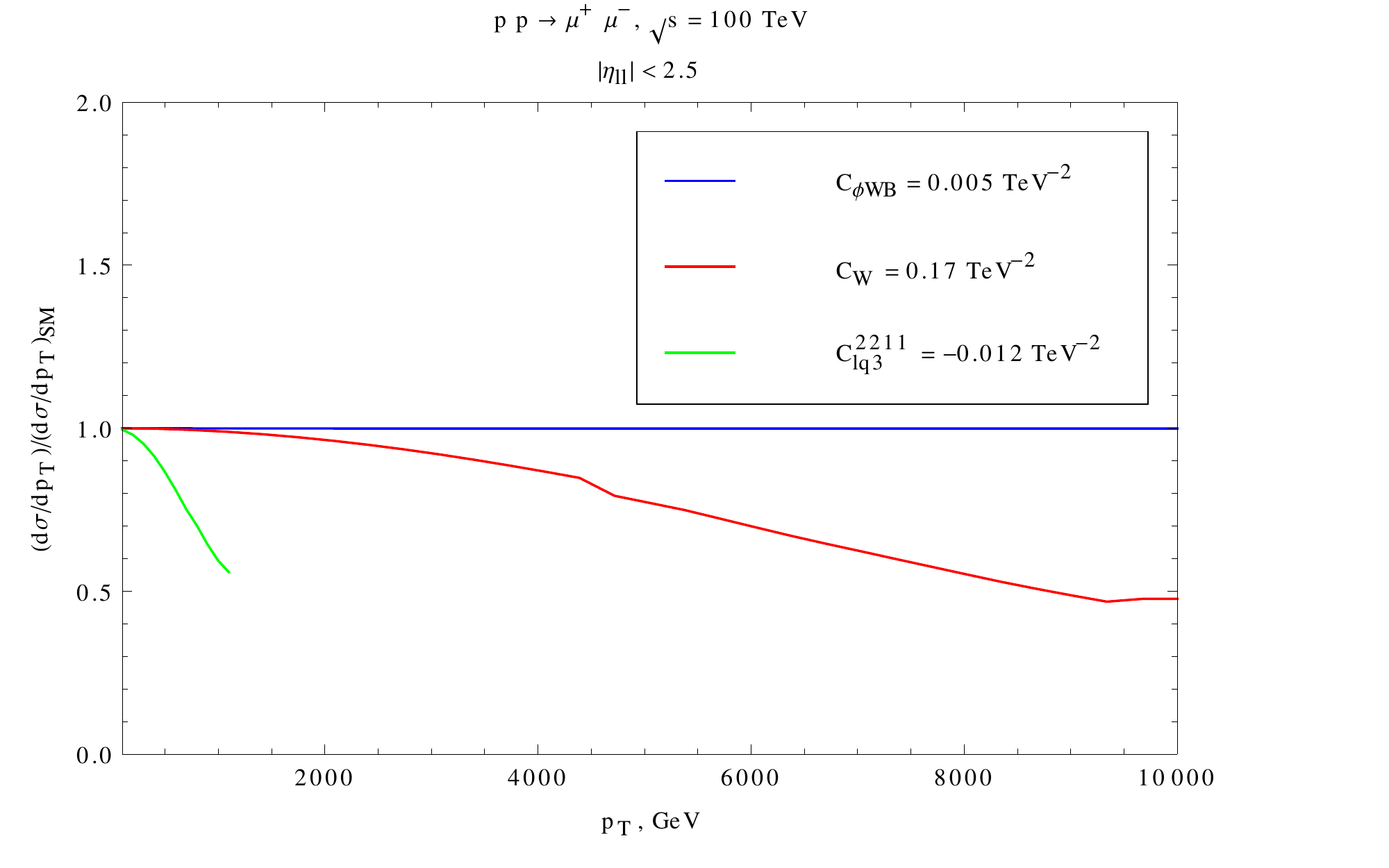}
\caption{Same as Fig.~\ref{fig:dsdmll}, but for the $p_T$ distribution.}
\label{fig:dsdpt}
\end{figure}

In Figure~\ref{fig:dsdmll}, we show the fractional modification of the dilepton mass distribution due to different bosonic operators. We include not only the operators $\Op_{\vp WB}$ and $\Op_{W}$, which cause loop effects and are the focus of our calculation, but also one of the 4-fermion operators that affects Drell-Yan production at tree level, for comparison. 
$\C_{\vp WB}$  corresponds to the $S$ parameter $S$~\cite{Peskin:1990zt,Peskin:1991sw} and is constrained to be
\begin{equation}
-0.004 < \C_{\vp WB}\biggl({1~\mathrm{TeV}\over\Lambda}\biggr)^2  < 0.006 \, ,
\label{eq:chwblimit}
\end{equation}
where we take our limit from the one-parameter fit of the  Gfitter collaboration~\cite{Haller:2018nnx}. The bound from the LEP precision electroweak data 
dominates the global fits of Refs.~\cite{Alves:2018nof,Biekotter:2018rhp,Almeida:2018cld,Bjorn:2016zlr}.  $C_{W}$ is bounded by $W^+W^-$ production, as described in Sec.~\ref{sec:nlo}. The 4-fermion operators are limited by existing Drell-Yan measurements~\cite{Greljo:2017vvb}, and of the many potential operators which contribute, the best constrained operator is $\underset{2,2,1,1}{\Op_{l q}^{(3)}}$, whose coefficient is limited by a one-parameter fit to be 
\begin{equation}
\label{eq:clq3limit}
-0.012 < \underset{2,2,1,1}{\C_{l q}^{(3)}}  \biggl({1~\mathrm{TeV}\over\Lambda}\biggr)^2< 0.0047\qquad({\text{Single~parameter~fit\cite{Greljo:2017vvb})}}\, .
\end{equation}
The results of the   global SMEFT fit of Ref. \cite{Bjorn:2016zlr} find considerably less stringent bounds on $C_{lq}^{(3)}$ due to large correlations between the
effects of different operators.

The left panel shows the effects of these operators at the LHC, while the right panel shows them at a future 100 TeV collider. For the 4-fermion operator, the EFT loses validity at high invariant mass because the neglected dimension-8 operators become important, and it is no longer appropriate to treat the EFT contribution as a small correction to the SM amplitude. We have cut off the associated curves where $A_{LL,q}^{4-fermions} = A_{LL}^{SM}/2$, which occurs at the center of mass energy $m_{\ell\ell}^{*} = 2.2$~TeV. Nevertheless, it is clear that in a non-universal theory, new 4-fermion operators can change Drell-Yan production much more than purely bosonic operators, even those such as $\C_{\vp WB}$ which contribute at tree level. At 100 TeV, where Drell-Yan could be potentially measured up to $m_{\ell \ell} = 20~\mathrm{TeV}$~\cite{Mangano:2016jyj}, the operator $\C_{W}$ at its current $2\sigma$ limit could provide up to a 50\% deviation in the number of events at high energy. The effects are similarly sized for the $p_T$ distribution, as shown in Figure~\ref{fig:dsdpt}. Again, the 4-fermion operator contributions become large at high $p_T$, and we avoid any phase space corresponding to an invariant mass higher than $m_{\ell\ell}^{*}$ by cutting off the corresponding curves at $p_T = m_{\ell \ell}^{*} / 2 = 1.1$~TeV.

\begin{figure}
\centering
\includegraphics[width=0.49\textwidth]{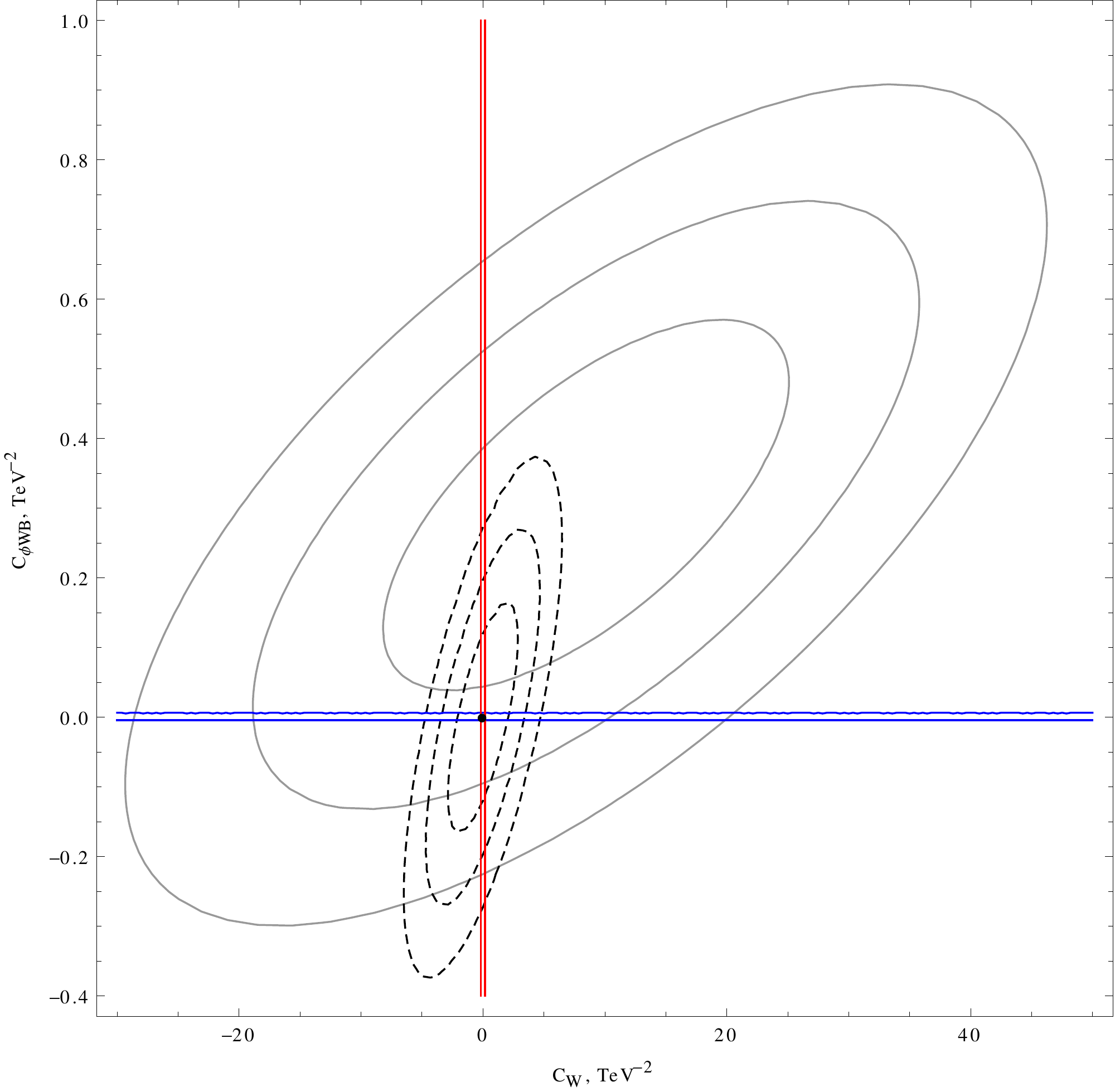}
\caption{Contours in the $\C_{W}-\C_{\vp WB}$ plane resulting from a fit to the CMS measurement of Drell-Yan. The contours indicate 68.3\%, 95.4\%, and 99.7\% confidence contours around the best-fit point. Solid contours are for the current 8 TeV measurement~\cite{CMS:2014jea}, while dashed contours are for the HL-LHC. The Standard Model is indicated at (0, 0). The region between the blue (red) lines is allowed by the current limits of Eq.~\ref{eq:chwblimit} (Eq.~\ref{eq:cwlimit}).}
\label{fig:cmsdy}
\end{figure}

From our derived effects of new operators on the Drell-Yan $m_{\ell\ell}$ distribution, we can place limits on the sizes of these operators from existing measurements of Drell-Yan production. We use the 8 TeV CMS measurement~\cite{CMS:2014jea}, which goes up to 2 TeV in the dilepton invariant mass. Using the CMS data above 240 GeV and neglecting correlated uncertainties among different bins\footnote{At high invariant mass, in the region where the SMEFT operators are expected to have the greatest effect, the uncertainty is dominantly statistical.}, we construct a $\chi^2$ function expressing the goodness of fit between the observed data and the prediction for arbitrary sizes of the SMEFT operators $\Op_{W}$ and $\Op_{\vp WB}$. We consider only how the new operators affect the ratio of the data to theory, and are not sensitive to the overall normalization of the Drell-Yan invariant mass distribution. In Figure~\ref{fig:cmsdy}, we show the allowed region in the plane of the sizes of these two operators. The 13 TeV measurement~\cite{CMS-PAS-SMP-16-019} using 2.8~fb$^{-1}$ of data already has comparable uncertainties in bins going out to 3 TeV in $m_{\ell\ell}$, and we also show a projection for the high luminosity upgrade of the LHC assuming that statistical uncertainties scale as $1/\sqrt{L}$ such that the uncertainties in each bin would be limited only by systematics, which are currently around $5\%$. 
The currently allowed region in the plane is significantly larger than that allowed by constraints from electroweak precision and $WW$ production on the operator coefficients $\C_{\vp WB}$ and $\C_{W}$, respectively. However, as both $\Op_{\vp WB}$ and $\Op_{W}$ contribute to Drell-Yan, an external constraint on one of the operator coefficients in conjunction with a measurement of the Drell-Yan differential distribution can constrain the other better.

\section{Conclusion}
\label{sec:concl}

In the absence of new light physics, the EFT approach provides a parametrization of BSM effects in terms of higher dimension operators. Given the minute precision with which many processes can be measured at the LHC, especially at high luminosity, it is of use to know the loop contributions of EFT operators to SM physics. In this work, we have evaluated some of the these corrections in the SMEFT for Drell-Yan production.

Notably, operators which do not contribute at tree level to Drell-Yan can have sizable contributions at one loop. We have shown that at the upper end of the energy range that can be probed at the LHC, the effect of the operator $\Op_W$ can be several percent, or even up to 50\% at a future 100 TeV collider. Over the lifetime of the LHC, statistical uncertainties will go down significantly as more data is collected, significantly increasing the sensitivity of precision Drell-Yan measurements in the region where new physics operators have the greatest effect. As a consequence, future measurements of Drell-Yan offer the possibility to constrain operators such as $\Op_W$ even though their contributions are only at one loop. While gauge boson production is still a more sensitive probe of $\Op_W$, the effects which we have computed here must be taken into account to ensure consistency in a full NLO fit to the coefficients of SMEFT operators.

\section*{Acknowledgements}SD is supported by the U.S.~Department of Energy under Grant Contract DE-SC0012704. PPG is supported by the Spanish Research Agency (Agencia Estatal de Investigacion) through the contract FPA2016-78022-P and IFT Centro de Excelencia Severo Ochoa under grant SEV-2016-0597.  AI is supported by the U.S.~Department of Energy under Grant Contract DE-SC0015634 and by PITT PACC.
Digital data related to our results can
be found
at~\url{https://quark.phy.bnl.gov/Digital_Data_Archive/dawson/drellyan_18}.

\bibliography{dy}

\begin{thebibliography}{56}
\expandafter\ifx\csname natexlab\endcsname\relax\def\natexlab#1{#1}\fi
\expandafter\ifx\csname bibnamefont\endcsname\relax
  \def\bibnamefont#1{#1}\fi
\expandafter\ifx\csname bibfnamefont\endcsname\relax
  \def\bibfnamefont#1{#1}\fi
\expandafter\ifx\csname citenamefont\endcsname\relax
  \def\citenamefont#1{#1}\fi
\expandafter\ifx\csname url\endcsname\relax
  \def\url#1{\texttt{#1}}\fi
\expandafter\ifx\csname urlprefix\endcsname\relax\def\urlprefix{URL }\fi
\providecommand{\bibinfo}[2]{#2}
\providecommand{\eprint}[2][]{\url{#2}}

\bibitem[{\citenamefont{Giudice et~al.}(2007)\citenamefont{Giudice, Grojean,
  Pomarol, and Rattazzi}}]{Giudice:2007fh}
\bibinfo{author}{\bibfnamefont{G.~F.} \bibnamefont{Giudice}},
  \bibinfo{author}{\bibfnamefont{C.}~\bibnamefont{Grojean}},
  \bibinfo{author}{\bibfnamefont{A.}~\bibnamefont{Pomarol}}, \bibnamefont{and}
  \bibinfo{author}{\bibfnamefont{R.}~\bibnamefont{Rattazzi}},
  \bibinfo{journal}{JHEP} \textbf{\bibinfo{volume}{06}}, \bibinfo{pages}{045}
  (\bibinfo{year}{2007}), \eprint{hep-ph/0703164}.

\bibitem[{\citenamefont{Brivio and Trott}(2017{\natexlab{a}})}]{Brivio:2017vri}
\bibinfo{author}{\bibfnamefont{I.}~\bibnamefont{Brivio}} \bibnamefont{and}
  \bibinfo{author}{\bibfnamefont{M.}~\bibnamefont{Trott}}
  (\bibinfo{year}{2017}{\natexlab{a}}), \eprint{1706.08945}.

\bibitem[{\citenamefont{Baglio et~al.}(2017)\citenamefont{Baglio, Dawson, and
  Lewis}}]{Baglio:2017bfe}
\bibinfo{author}{\bibfnamefont{J.}~\bibnamefont{Baglio}},
  \bibinfo{author}{\bibfnamefont{S.}~\bibnamefont{Dawson}}, \bibnamefont{and}
  \bibinfo{author}{\bibfnamefont{I.~M.} \bibnamefont{Lewis}},
  \bibinfo{journal}{Phys. Rev.} \textbf{\bibinfo{volume}{D96}},
  \bibinfo{pages}{073003} (\bibinfo{year}{2017}), \eprint{1708.03332}.

\bibitem[{\citenamefont{Azatov et~al.}(2017)\citenamefont{Azatov, Contino,
  Machado, and Riva}}]{Azatov:2016sqh}
\bibinfo{author}{\bibfnamefont{A.}~\bibnamefont{Azatov}},
  \bibinfo{author}{\bibfnamefont{R.}~\bibnamefont{Contino}},
  \bibinfo{author}{\bibfnamefont{C.~S.} \bibnamefont{Machado}},
  \bibnamefont{and} \bibinfo{author}{\bibfnamefont{F.}~\bibnamefont{Riva}},
  \bibinfo{journal}{Phys. Rev.} \textbf{\bibinfo{volume}{D95}},
  \bibinfo{pages}{065014} (\bibinfo{year}{2017}), \eprint{1607.05236}.

\bibitem[{\citenamefont{Zhang}(2017)}]{Zhang:2016zsp}
\bibinfo{author}{\bibfnamefont{Z.}~\bibnamefont{Zhang}},
  \bibinfo{journal}{Phys. Rev. Lett.} \textbf{\bibinfo{volume}{118}},
  \bibinfo{pages}{011803} (\bibinfo{year}{2017}), \eprint{1610.01618}.

\bibitem[{\citenamefont{Grojean et~al.}(2018)\citenamefont{Grojean, Montull,
  and Riembau}}]{Grojean:2018dqj}
\bibinfo{author}{\bibfnamefont{C.}~\bibnamefont{Grojean}},
  \bibinfo{author}{\bibfnamefont{M.}~\bibnamefont{Montull}}, \bibnamefont{and}
  \bibinfo{author}{\bibfnamefont{M.}~\bibnamefont{Riembau}}
  (\bibinfo{year}{2018}), \eprint{1810.05149}.

\bibitem[{\citenamefont{Alves et~al.}(2018)\citenamefont{Alves, Rosa-Agostinho,
  Eboli, and Gonzalez-Garcia}}]{Alves:2018nof}
\bibinfo{author}{\bibfnamefont{A.}~\bibnamefont{Alves}},
  \bibinfo{author}{\bibfnamefont{N.}~\bibnamefont{Rosa-Agostinho}},
  \bibinfo{author}{\bibfnamefont{O.~J.~P.} \bibnamefont{Eboli}},
  \bibnamefont{and} \bibinfo{author}{\bibfnamefont{M.~C.}
  \bibnamefont{Gonzalez-Garcia}}, \bibinfo{journal}{Phys. Rev.}
  \textbf{\bibinfo{volume}{D98}}, \bibinfo{pages}{013006}
  (\bibinfo{year}{2018}), \eprint{1805.11108}.

\bibitem[{\citenamefont{Barducci et~al.}(2018)}]{AguilarSaavedra:2018nen}
\bibinfo{author}{\bibfnamefont{D.}~\bibnamefont{Barducci}} \bibnamefont{et~al.}
  (\bibinfo{year}{2018}), \eprint{1802.07237}.

\bibitem[{\citenamefont{Degrande et~al.}(2012)\citenamefont{Degrande, Gerard,
  Grojean, Maltoni, and Servant}}]{Degrande:2012gr}
\bibinfo{author}{\bibfnamefont{C.}~\bibnamefont{Degrande}},
  \bibinfo{author}{\bibfnamefont{J.~M.} \bibnamefont{Gerard}},
  \bibinfo{author}{\bibfnamefont{C.}~\bibnamefont{Grojean}},
  \bibinfo{author}{\bibfnamefont{F.}~\bibnamefont{Maltoni}}, \bibnamefont{and}
  \bibinfo{author}{\bibfnamefont{G.}~\bibnamefont{Servant}},
  \bibinfo{journal}{JHEP} \textbf{\bibinfo{volume}{07}}, \bibinfo{pages}{036}
  (\bibinfo{year}{2012}), \bibinfo{note}{[Erratum: JHEP03,032(2013)]},
  \eprint{1205.1065}.

\bibitem[{\citenamefont{Farina et~al.}(2018)\citenamefont{Farina, Mondino,
  Pappadopulo, and Ruderman}}]{Farina:2018lqo}
\bibinfo{author}{\bibfnamefont{M.}~\bibnamefont{Farina}},
  \bibinfo{author}{\bibfnamefont{C.}~\bibnamefont{Mondino}},
  \bibinfo{author}{\bibfnamefont{D.}~\bibnamefont{Pappadopulo}},
  \bibnamefont{and} \bibinfo{author}{\bibfnamefont{J.~T.}
  \bibnamefont{Ruderman}} (\bibinfo{year}{2018}), \eprint{1811.04084}.

\bibitem[{\citenamefont{Anastasiou et~al.}(2003)\citenamefont{Anastasiou,
  Dixon, Melnikov, and Petriello}}]{Anastasiou:2003yy}
\bibinfo{author}{\bibfnamefont{C.}~\bibnamefont{Anastasiou}},
  \bibinfo{author}{\bibfnamefont{L.~J.} \bibnamefont{Dixon}},
  \bibinfo{author}{\bibfnamefont{K.}~\bibnamefont{Melnikov}}, \bibnamefont{and}
  \bibinfo{author}{\bibfnamefont{F.}~\bibnamefont{Petriello}},
  \bibinfo{journal}{Phys. Rev. Lett.} \textbf{\bibinfo{volume}{91}},
  \bibinfo{pages}{182002} (\bibinfo{year}{2003}), \eprint{hep-ph/0306192}.

\bibitem[{\citenamefont{Gehrmann-De~Ridder
  et~al.}(2016{\natexlab{a}})\citenamefont{Gehrmann-De~Ridder, Gehrmann,
  Glover, Huss, and Morgan}}]{Gehrmann-DeRidder:2016jns}
\bibinfo{author}{\bibfnamefont{A.}~\bibnamefont{Gehrmann-De~Ridder}},
  \bibinfo{author}{\bibfnamefont{T.}~\bibnamefont{Gehrmann}},
  \bibinfo{author}{\bibfnamefont{E.~W.~N.} \bibnamefont{Glover}},
  \bibinfo{author}{\bibfnamefont{A.}~\bibnamefont{Huss}}, \bibnamefont{and}
  \bibinfo{author}{\bibfnamefont{T.~A.} \bibnamefont{Morgan}},
  \bibinfo{journal}{JHEP} \textbf{\bibinfo{volume}{11}}, \bibinfo{pages}{094}
  (\bibinfo{year}{2016}{\natexlab{a}}), \bibinfo{note}{[Erratum:
  JHEP10,126(2018)]}, \eprint{1610.01843}.

\bibitem[{\citenamefont{Gehrmann-De~Ridder
  et~al.}(2016{\natexlab{b}})\citenamefont{Gehrmann-De~Ridder, Gehrmann,
  Glover, Huss, and Morgan}}]{Ridder:2016nkl}
\bibinfo{author}{\bibfnamefont{A.}~\bibnamefont{Gehrmann-De~Ridder}},
  \bibinfo{author}{\bibfnamefont{T.}~\bibnamefont{Gehrmann}},
  \bibinfo{author}{\bibfnamefont{E.~W.~N.} \bibnamefont{Glover}},
  \bibinfo{author}{\bibfnamefont{A.}~\bibnamefont{Huss}}, \bibnamefont{and}
  \bibinfo{author}{\bibfnamefont{T.~A.} \bibnamefont{Morgan}},
  \bibinfo{journal}{JHEP} \textbf{\bibinfo{volume}{07}}, \bibinfo{pages}{133}
  (\bibinfo{year}{2016}{\natexlab{b}}), \eprint{1605.04295}.

\bibitem[{\citenamefont{Karlberg et~al.}(2014)\citenamefont{Karlberg, Re, and
  Zanderighi}}]{Karlberg:2014qua}
\bibinfo{author}{\bibfnamefont{A.}~\bibnamefont{Karlberg}},
  \bibinfo{author}{\bibfnamefont{E.}~\bibnamefont{Re}}, \bibnamefont{and}
  \bibinfo{author}{\bibfnamefont{G.}~\bibnamefont{Zanderighi}},
  \bibinfo{journal}{JHEP} \textbf{\bibinfo{volume}{09}}, \bibinfo{pages}{134}
  (\bibinfo{year}{2014}), \eprint{1407.2940}.

\bibitem[{\citenamefont{Hoche et~al.}(2015)\citenamefont{Hoche, Li, and
  Prestel}}]{Hoeche:2014aia}
\bibinfo{author}{\bibfnamefont{S.}~\bibnamefont{Hoche}},
  \bibinfo{author}{\bibfnamefont{Y.}~\bibnamefont{Li}}, \bibnamefont{and}
  \bibinfo{author}{\bibfnamefont{S.}~\bibnamefont{Prestel}},
  \bibinfo{journal}{Phys. Rev.} \textbf{\bibinfo{volume}{D91}},
  \bibinfo{pages}{074015} (\bibinfo{year}{2015}), \eprint{1405.3607}.

\bibitem[{\citenamefont{Bizo et~al.}(2018)\citenamefont{Bizo, Chen,
  Gehrmann-De~Ridder, Gehrmann, Glover, Huss, Monni, Re, Rottoli, and
  Torrielli}}]{Bizon:2018foh}
\bibinfo{author}{\bibfnamefont{W.}~\bibnamefont{Bizo}},
  \bibinfo{author}{\bibfnamefont{X.}~\bibnamefont{Chen}},
  \bibinfo{author}{\bibfnamefont{A.}~\bibnamefont{Gehrmann-De~Ridder}},
  \bibinfo{author}{\bibfnamefont{T.}~\bibnamefont{Gehrmann}},
  \bibinfo{author}{\bibfnamefont{N.}~\bibnamefont{Glover}},
  \bibinfo{author}{\bibfnamefont{A.}~\bibnamefont{Huss}},
  \bibinfo{author}{\bibfnamefont{P.~F.} \bibnamefont{Monni}},
  \bibinfo{author}{\bibfnamefont{E.}~\bibnamefont{Re}},
  \bibinfo{author}{\bibfnamefont{L.}~\bibnamefont{Rottoli}}, \bibnamefont{and}
  \bibinfo{author}{\bibfnamefont{P.}~\bibnamefont{Torrielli}}
  (\bibinfo{year}{2018}), \eprint{1805.05916}.

\bibitem[{\citenamefont{Alioli et~al.}(2015)\citenamefont{Alioli, Bauer,
  Berggren, Tackmann, and Walsh}}]{Alioli:2015toa}
\bibinfo{author}{\bibfnamefont{S.}~\bibnamefont{Alioli}},
  \bibinfo{author}{\bibfnamefont{C.~W.} \bibnamefont{Bauer}},
  \bibinfo{author}{\bibfnamefont{C.}~\bibnamefont{Berggren}},
  \bibinfo{author}{\bibfnamefont{F.~J.} \bibnamefont{Tackmann}},
  \bibnamefont{and} \bibinfo{author}{\bibfnamefont{J.~R.} \bibnamefont{Walsh}},
  \bibinfo{journal}{Phys. Rev.} \textbf{\bibinfo{volume}{D92}},
  \bibinfo{pages}{094020} (\bibinfo{year}{2015}), \eprint{1508.01475}.

\bibitem[{\citenamefont{Dittmaier and Huber}(2010)}]{Dittmaier:2009cr}
\bibinfo{author}{\bibfnamefont{S.}~\bibnamefont{Dittmaier}} \bibnamefont{and}
  \bibinfo{author}{\bibfnamefont{M.}~\bibnamefont{Huber}},
  \bibinfo{journal}{JHEP} \textbf{\bibinfo{volume}{01}}, \bibinfo{pages}{060}
  (\bibinfo{year}{2010}), \eprint{0911.2329}.

\bibitem[{\citenamefont{Baur et~al.}(2002)\citenamefont{Baur, Brein, Hollik,
  Schappacher, and Wackeroth}}]{Baur:2001ze}
\bibinfo{author}{\bibfnamefont{U.}~\bibnamefont{Baur}},
  \bibinfo{author}{\bibfnamefont{O.}~\bibnamefont{Brein}},
  \bibinfo{author}{\bibfnamefont{W.}~\bibnamefont{Hollik}},
  \bibinfo{author}{\bibfnamefont{C.}~\bibnamefont{Schappacher}},
  \bibnamefont{and}
  \bibinfo{author}{\bibfnamefont{D.}~\bibnamefont{Wackeroth}},
  \bibinfo{journal}{Phys. Rev.} \textbf{\bibinfo{volume}{D65}},
  \bibinfo{pages}{033007} (\bibinfo{year}{2002}), \eprint{hep-ph/0108274}.

\bibitem[{\citenamefont{Li and Petriello}(2012)}]{Li:2012wna}
\bibinfo{author}{\bibfnamefont{Y.}~\bibnamefont{Li}} \bibnamefont{and}
  \bibinfo{author}{\bibfnamefont{F.}~\bibnamefont{Petriello}},
  \bibinfo{journal}{Phys. Rev.} \textbf{\bibinfo{volume}{D86}},
  \bibinfo{pages}{094034} (\bibinfo{year}{2012}), \eprint{1208.5967}.

\bibitem[{\citenamefont{Gavin et~al.}(2011)\citenamefont{Gavin, Li, Petriello,
  and Quackenbush}}]{Gavin:2010az}
\bibinfo{author}{\bibfnamefont{R.}~\bibnamefont{Gavin}},
  \bibinfo{author}{\bibfnamefont{Y.}~\bibnamefont{Li}},
  \bibinfo{author}{\bibfnamefont{F.}~\bibnamefont{Petriello}},
  \bibnamefont{and}
  \bibinfo{author}{\bibfnamefont{S.}~\bibnamefont{Quackenbush}},
  \bibinfo{journal}{Comput. Phys. Commun.} \textbf{\bibinfo{volume}{182}},
  \bibinfo{pages}{2388} (\bibinfo{year}{2011}), \eprint{1011.3540}.

\bibitem[{\citenamefont{Carloni~Calame
  et~al.}(2007)\citenamefont{Carloni~Calame, Montagna, Nicrosini, and
  Vicini}}]{CarloniCalame:2007cd}
\bibinfo{author}{\bibfnamefont{C.~M.} \bibnamefont{Carloni~Calame}},
  \bibinfo{author}{\bibfnamefont{G.}~\bibnamefont{Montagna}},
  \bibinfo{author}{\bibfnamefont{O.}~\bibnamefont{Nicrosini}},
  \bibnamefont{and} \bibinfo{author}{\bibfnamefont{A.}~\bibnamefont{Vicini}},
  \bibinfo{journal}{JHEP} \textbf{\bibinfo{volume}{10}}, \bibinfo{pages}{109}
  (\bibinfo{year}{2007}), \eprint{0710.1722}.

\bibitem[{\citenamefont{Grzadkowski et~al.}(2010)\citenamefont{Grzadkowski,
  Iskrzynski, Misiak, and Rosiek}}]{Grzadkowski:2010es}
\bibinfo{author}{\bibfnamefont{B.}~\bibnamefont{Grzadkowski}},
  \bibinfo{author}{\bibfnamefont{M.}~\bibnamefont{Iskrzynski}},
  \bibinfo{author}{\bibfnamefont{M.}~\bibnamefont{Misiak}}, \bibnamefont{and}
  \bibinfo{author}{\bibfnamefont{J.}~\bibnamefont{Rosiek}},
  \bibinfo{journal}{JHEP} \textbf{\bibinfo{volume}{10}}, \bibinfo{pages}{085}
  (\bibinfo{year}{2010}), \eprint{1008.4884}.

\bibitem[{\citenamefont{Falkowski et~al.}(2017)\citenamefont{Falkowski,
  Gonz\'{a}lez-Alonso, and Mimouni}}]{Falkowski:2017pss}
\bibinfo{author}{\bibfnamefont{A.}~\bibnamefont{Falkowski}},
  \bibinfo{author}{\bibfnamefont{M.}~\bibnamefont{Gonz\'{a}lez-Alonso}},
  \bibnamefont{and} \bibinfo{author}{\bibfnamefont{K.}~\bibnamefont{Mimouni}},
  \bibinfo{journal}{JHEP} \textbf{\bibinfo{volume}{08}}, \bibinfo{pages}{123}
  (\bibinfo{year}{2017}), \eprint{1706.03783}.

\bibitem[{\citenamefont{Berthier and Trott}(2015)}]{Berthier:2015oma}
\bibinfo{author}{\bibfnamefont{L.}~\bibnamefont{Berthier}} \bibnamefont{and}
  \bibinfo{author}{\bibfnamefont{M.}~\bibnamefont{Trott}},
  \bibinfo{journal}{JHEP} \textbf{\bibinfo{volume}{05}}, \bibinfo{pages}{024}
  (\bibinfo{year}{2015}), \eprint{1502.02570}.

\bibitem[{\citenamefont{de~Blas et~al.}(2013)\citenamefont{de~Blas, Chala, and
  Santiago}}]{deBlas:2013qqa}
\bibinfo{author}{\bibfnamefont{J.}~\bibnamefont{de~Blas}},
  \bibinfo{author}{\bibfnamefont{M.}~\bibnamefont{Chala}}, \bibnamefont{and}
  \bibinfo{author}{\bibfnamefont{J.}~\bibnamefont{Santiago}},
  \bibinfo{journal}{Phys. Rev.} \textbf{\bibinfo{volume}{D88}},
  \bibinfo{pages}{095011} (\bibinfo{year}{2013}), \eprint{1307.5068}.

\bibitem[{\citenamefont{Cirigliano et~al.}(2013)\citenamefont{Cirigliano,
  Gonzalez-Alonso, and Graesser}}]{Cirigliano:2012ab}
\bibinfo{author}{\bibfnamefont{V.}~\bibnamefont{Cirigliano}},
  \bibinfo{author}{\bibfnamefont{M.}~\bibnamefont{Gonzalez-Alonso}},
  \bibnamefont{and} \bibinfo{author}{\bibfnamefont{M.~L.}
  \bibnamefont{Graesser}}, \bibinfo{journal}{JHEP}
  \textbf{\bibinfo{volume}{02}}, \bibinfo{pages}{046} (\bibinfo{year}{2013}),
  \eprint{1210.4553}.

\bibitem[{\citenamefont{Carpentier and Davidson}(2010)}]{Carpentier:2010ue}
\bibinfo{author}{\bibfnamefont{M.}~\bibnamefont{Carpentier}} \bibnamefont{and}
  \bibinfo{author}{\bibfnamefont{S.}~\bibnamefont{Davidson}},
  \bibinfo{journal}{Eur. Phys. J.} \textbf{\bibinfo{volume}{C70}},
  \bibinfo{pages}{1071} (\bibinfo{year}{2010}), \eprint{1008.0280}.

\bibitem[{\citenamefont{Alioli et~al.}(2018)\citenamefont{Alioli, Dekens,
  Girard, and Mereghetti}}]{Alioli:2018ljm}
\bibinfo{author}{\bibfnamefont{S.}~\bibnamefont{Alioli}},
  \bibinfo{author}{\bibfnamefont{W.}~\bibnamefont{Dekens}},
  \bibinfo{author}{\bibfnamefont{M.}~\bibnamefont{Girard}}, \bibnamefont{and}
  \bibinfo{author}{\bibfnamefont{E.}~\bibnamefont{Mereghetti}},
  \bibinfo{journal}{JHEP} \textbf{\bibinfo{volume}{08}}, \bibinfo{pages}{205}
  (\bibinfo{year}{2018}), \eprint{1804.07407}.

\bibitem[{\citenamefont{Dawson and
  Giardino}(2018{\natexlab{a}})}]{Dawson:2018pyl}
\bibinfo{author}{\bibfnamefont{S.}~\bibnamefont{Dawson}} \bibnamefont{and}
  \bibinfo{author}{\bibfnamefont{P.~P.} \bibnamefont{Giardino}},
  \bibinfo{journal}{Phys. Rev.} \textbf{\bibinfo{volume}{D97}},
  \bibinfo{pages}{093003} (\bibinfo{year}{2018}{\natexlab{a}}),
  \eprint{1801.01136}.

\bibitem[{\citenamefont{Dawson and
  Giardino}(2018{\natexlab{b}})}]{Dawson:2018liq}
\bibinfo{author}{\bibfnamefont{S.}~\bibnamefont{Dawson}} \bibnamefont{and}
  \bibinfo{author}{\bibfnamefont{P.~P.} \bibnamefont{Giardino}}
  (\bibinfo{year}{2018}{\natexlab{b}}), \eprint{1807.11504}.

\bibitem[{\citenamefont{Hartmann and Trott}(2015)}]{Hartmann:2015aia}
\bibinfo{author}{\bibfnamefont{C.}~\bibnamefont{Hartmann}} \bibnamefont{and}
  \bibinfo{author}{\bibfnamefont{M.}~\bibnamefont{Trott}},
  \bibinfo{journal}{Phys. Rev. Lett.} \textbf{\bibinfo{volume}{115}},
  \bibinfo{pages}{191801} (\bibinfo{year}{2015}), \eprint{1507.03568}.

\bibitem[{\citenamefont{Dedes et~al.}(2018)\citenamefont{Dedes, Paraskevas,
  Rosiek, Suxho, and Trifyllis}}]{Dedes:2018seb}
\bibinfo{author}{\bibfnamefont{A.}~\bibnamefont{Dedes}},
  \bibinfo{author}{\bibfnamefont{M.}~\bibnamefont{Paraskevas}},
  \bibinfo{author}{\bibfnamefont{J.}~\bibnamefont{Rosiek}},
  \bibinfo{author}{\bibfnamefont{K.}~\bibnamefont{Suxho}}, \bibnamefont{and}
  \bibinfo{author}{\bibfnamefont{L.}~\bibnamefont{Trifyllis}},
  \bibinfo{journal}{JHEP} \textbf{\bibinfo{volume}{08}}, \bibinfo{pages}{103}
  (\bibinfo{year}{2018}), \eprint{1805.00302}.

\bibitem[{\citenamefont{Gauld et~al.}(2016{\natexlab{a}})\citenamefont{Gauld,
  Pecjak, and Scott}}]{Gauld:2016kuu}
\bibinfo{author}{\bibfnamefont{R.}~\bibnamefont{Gauld}},
  \bibinfo{author}{\bibfnamefont{B.~D.} \bibnamefont{Pecjak}},
  \bibnamefont{and} \bibinfo{author}{\bibfnamefont{D.~J.} \bibnamefont{Scott}},
  \bibinfo{journal}{Phys. Rev.} \textbf{\bibinfo{volume}{D94}},
  \bibinfo{pages}{074045} (\bibinfo{year}{2016}{\natexlab{a}}),
  \eprint{1607.06354}.

\bibitem[{\citenamefont{Gauld et~al.}(2016{\natexlab{b}})\citenamefont{Gauld,
  Pecjak, and Scott}}]{Gauld:2015lmb}
\bibinfo{author}{\bibfnamefont{R.}~\bibnamefont{Gauld}},
  \bibinfo{author}{\bibfnamefont{B.~D.} \bibnamefont{Pecjak}},
  \bibnamefont{and} \bibinfo{author}{\bibfnamefont{D.~J.} \bibnamefont{Scott}},
  \bibinfo{journal}{JHEP} \textbf{\bibinfo{volume}{05}}, \bibinfo{pages}{080}
  (\bibinfo{year}{2016}{\natexlab{b}}), \eprint{1512.02508}.

\bibitem[{\citenamefont{Hartmann et~al.}(2017)\citenamefont{Hartmann, Shepherd,
  and Trott}}]{Hartmann:2016pil}
\bibinfo{author}{\bibfnamefont{C.}~\bibnamefont{Hartmann}},
  \bibinfo{author}{\bibfnamefont{W.}~\bibnamefont{Shepherd}}, \bibnamefont{and}
  \bibinfo{author}{\bibfnamefont{M.}~\bibnamefont{Trott}},
  \bibinfo{journal}{JHEP} \textbf{\bibinfo{volume}{03}}, \bibinfo{pages}{060}
  (\bibinfo{year}{2017}), \eprint{1611.09879}.

\bibitem[{\citenamefont{Dawson and Ismail}(2018)}]{Dawson:2018jlg}
\bibinfo{author}{\bibfnamefont{S.}~\bibnamefont{Dawson}} \bibnamefont{and}
  \bibinfo{author}{\bibfnamefont{A.}~\bibnamefont{Ismail}},
  \bibinfo{journal}{Phys. Rev.} \textbf{\bibinfo{volume}{D98}},
  \bibinfo{pages}{093003} (\bibinfo{year}{2018}), \eprint{1808.05948}.

\bibitem[{\citenamefont{Farina et~al.}(2017)\citenamefont{Farina, Panico,
  Pappadopulo, Ruderman, Torre, and Wulzer}}]{Farina:2016rws}
\bibinfo{author}{\bibfnamefont{M.}~\bibnamefont{Farina}},
  \bibinfo{author}{\bibfnamefont{G.}~\bibnamefont{Panico}},
  \bibinfo{author}{\bibfnamefont{D.}~\bibnamefont{Pappadopulo}},
  \bibinfo{author}{\bibfnamefont{J.~T.} \bibnamefont{Ruderman}},
  \bibinfo{author}{\bibfnamefont{R.}~\bibnamefont{Torre}}, \bibnamefont{and}
  \bibinfo{author}{\bibfnamefont{A.}~\bibnamefont{Wulzer}},
  \bibinfo{journal}{Phys. Lett.} \textbf{\bibinfo{volume}{B772}},
  \bibinfo{pages}{210} (\bibinfo{year}{2017}), \eprint{1609.08157}.

\bibitem[{\citenamefont{Buchmuller and Wyler}(1986)}]{Buchmuller:1985jz}
\bibinfo{author}{\bibfnamefont{W.}~\bibnamefont{Buchmuller}} \bibnamefont{and}
  \bibinfo{author}{\bibfnamefont{D.}~\bibnamefont{Wyler}},
  \bibinfo{journal}{Nucl. Phys.} \textbf{\bibinfo{volume}{B268}},
  \bibinfo{pages}{621} (\bibinfo{year}{1986}).

\bibitem[{\citenamefont{Hagiwara et~al.}(1993)\citenamefont{Hagiwara, Ishihara,
  Szalapski, and Zeppenfeld}}]{Hagiwara:1993ck}
\bibinfo{author}{\bibfnamefont{K.}~\bibnamefont{Hagiwara}},
  \bibinfo{author}{\bibfnamefont{S.}~\bibnamefont{Ishihara}},
  \bibinfo{author}{\bibfnamefont{R.}~\bibnamefont{Szalapski}},
  \bibnamefont{and}
  \bibinfo{author}{\bibfnamefont{D.}~\bibnamefont{Zeppenfeld}},
  \bibinfo{journal}{Phys. Rev.} \textbf{\bibinfo{volume}{D48}},
  \bibinfo{pages}{2182} (\bibinfo{year}{1993}).

\bibitem[{\citenamefont{Dedes et~al.}(2017)\citenamefont{Dedes, Materkowska,
  Paraskevas, Rosiek, and Suxho}}]{Dedes:2017zog}
\bibinfo{author}{\bibfnamefont{A.}~\bibnamefont{Dedes}},
  \bibinfo{author}{\bibfnamefont{W.}~\bibnamefont{Materkowska}},
  \bibinfo{author}{\bibfnamefont{M.}~\bibnamefont{Paraskevas}},
  \bibinfo{author}{\bibfnamefont{J.}~\bibnamefont{Rosiek}}, \bibnamefont{and}
  \bibinfo{author}{\bibfnamefont{K.}~\bibnamefont{Suxho}},
  \bibinfo{journal}{JHEP} \textbf{\bibinfo{volume}{06}}, \bibinfo{pages}{143}
  (\bibinfo{year}{2017}), \eprint{1704.03888}.

\bibitem[{\citenamefont{Falkowski}(2016)}]{Falkowski:2015fla}
\bibinfo{author}{\bibfnamefont{A.}~\bibnamefont{Falkowski}},
  \bibinfo{journal}{Pramana} \textbf{\bibinfo{volume}{87}}, \bibinfo{pages}{39}
  (\bibinfo{year}{2016}), \eprint{1505.00046}.

\bibitem[{\citenamefont{Brivio and Trott}(2017{\natexlab{b}})}]{Brivio:2017bnu}
\bibinfo{author}{\bibfnamefont{I.}~\bibnamefont{Brivio}} \bibnamefont{and}
  \bibinfo{author}{\bibfnamefont{M.}~\bibnamefont{Trott}},
  \bibinfo{journal}{JHEP} \textbf{\bibinfo{volume}{07}}, \bibinfo{pages}{148}
  (\bibinfo{year}{2017}{\natexlab{b}}), \bibinfo{note}{[Addendum:
  JHEP05,136(2018)]}, \eprint{1701.06424}.

\bibitem[{\citenamefont{Alonso et~al.}(2014)\citenamefont{Alonso, Jenkins,
  Manohar, and Trott}}]{Alonso:2013hga}
\bibinfo{author}{\bibfnamefont{R.}~\bibnamefont{Alonso}},
  \bibinfo{author}{\bibfnamefont{E.~E.} \bibnamefont{Jenkins}},
  \bibinfo{author}{\bibfnamefont{A.~V.} \bibnamefont{Manohar}},
  \bibnamefont{and} \bibinfo{author}{\bibfnamefont{M.}~\bibnamefont{Trott}},
  \bibinfo{journal}{JHEP} \textbf{\bibinfo{volume}{04}}, \bibinfo{pages}{159}
  (\bibinfo{year}{2014}), \eprint{1312.2014}.

\bibitem[{\citenamefont{Biekotter et~al.}(2018)\citenamefont{Biekotter,
  Corbett, and Plehn}}]{Biekotter:2018rhp}
\bibinfo{author}{\bibfnamefont{A.}~\bibnamefont{Biekotter}},
  \bibinfo{author}{\bibfnamefont{T.}~\bibnamefont{Corbett}}, \bibnamefont{and}
  \bibinfo{author}{\bibfnamefont{T.}~\bibnamefont{Plehn}}
  (\bibinfo{year}{2018}), \eprint{1812.07587}.

\bibitem[{\citenamefont{da~Silva~Almeida
  et~al.}(2018)\citenamefont{da~Silva~Almeida, Alves, Rosa~Agostinho, Eboli,
  and Gonzalez-Garcia}}]{Almeida:2018cld}
\bibinfo{author}{\bibfnamefont{E.}~\bibnamefont{da~Silva~Almeida}},
  \bibinfo{author}{\bibfnamefont{A.}~\bibnamefont{Alves}},
  \bibinfo{author}{\bibfnamefont{N.}~\bibnamefont{Rosa~Agostinho}},
  \bibinfo{author}{\bibfnamefont{O.~J.~P.} \bibnamefont{Eboli}},
  \bibnamefont{and} \bibinfo{author}{\bibfnamefont{M.~C.}
  \bibnamefont{Gonzalez-Garcia}} (\bibinfo{year}{2018}), \eprint{1812.01009}.

\bibitem[{\citenamefont{Bjorn and Trott}(2016)}]{Bjorn:2016zlr}
\bibinfo{author}{\bibfnamefont{M.}~\bibnamefont{Bjorn}} \bibnamefont{and}
  \bibinfo{author}{\bibfnamefont{M.}~\bibnamefont{Trott}},
  \bibinfo{journal}{Phys. Lett.} \textbf{\bibinfo{volume}{B762}},
  \bibinfo{pages}{426} (\bibinfo{year}{2016}), \eprint{1606.06502}.

\bibitem[{\citenamefont{Chen et~al.}(2014)\citenamefont{Chen, Dawson, and
  Zhang}}]{Chen:2013kfa}
\bibinfo{author}{\bibfnamefont{C.-Y.} \bibnamefont{Chen}},
  \bibinfo{author}{\bibfnamefont{S.}~\bibnamefont{Dawson}}, \bibnamefont{and}
  \bibinfo{author}{\bibfnamefont{C.}~\bibnamefont{Zhang}},
  \bibinfo{journal}{Phys. Rev.} \textbf{\bibinfo{volume}{D89}},
  \bibinfo{pages}{015016} (\bibinfo{year}{2014}), \eprint{1311.3107}.

\bibitem[{\citenamefont{Wells and Zhang}(2016)}]{Wells:2015cre}
\bibinfo{author}{\bibfnamefont{J.~D.} \bibnamefont{Wells}} \bibnamefont{and}
  \bibinfo{author}{\bibfnamefont{Z.}~\bibnamefont{Zhang}},
  \bibinfo{journal}{JHEP} \textbf{\bibinfo{volume}{06}}, \bibinfo{pages}{122}
  (\bibinfo{year}{2016}), \eprint{1512.03056}.

\bibitem[{\citenamefont{Peskin and Takeuchi}(1990)}]{Peskin:1990zt}
\bibinfo{author}{\bibfnamefont{M.~E.} \bibnamefont{Peskin}} \bibnamefont{and}
  \bibinfo{author}{\bibfnamefont{T.}~\bibnamefont{Takeuchi}},
  \bibinfo{journal}{Phys. Rev. Lett.} \textbf{\bibinfo{volume}{65}},
  \bibinfo{pages}{964} (\bibinfo{year}{1990}).

\bibitem[{\citenamefont{Peskin and Takeuchi}(1992)}]{Peskin:1991sw}
\bibinfo{author}{\bibfnamefont{M.~E.} \bibnamefont{Peskin}} \bibnamefont{and}
  \bibinfo{author}{\bibfnamefont{T.}~\bibnamefont{Takeuchi}},
  \bibinfo{journal}{Phys. Rev.} \textbf{\bibinfo{volume}{D46}},
  \bibinfo{pages}{381} (\bibinfo{year}{1992}).

\bibitem[{\citenamefont{Haller et~al.}(2018)\citenamefont{Haller, Hoecker,
  Kogler, M{\"o}nig, Peiffer, and Stelzer}}]{Haller:2018nnx}
\bibinfo{author}{\bibfnamefont{J.}~\bibnamefont{Haller}},
  \bibinfo{author}{\bibfnamefont{A.}~\bibnamefont{Hoecker}},
  \bibinfo{author}{\bibfnamefont{R.}~\bibnamefont{Kogler}},
  \bibinfo{author}{\bibfnamefont{K.}~\bibnamefont{M{\"o}nig}},
  \bibinfo{author}{\bibfnamefont{T.}~\bibnamefont{Peiffer}}, \bibnamefont{and}
  \bibinfo{author}{\bibfnamefont{J.}~\bibnamefont{Stelzer}},
  \bibinfo{journal}{Eur. Phys. J.} \textbf{\bibinfo{volume}{C78}},
  \bibinfo{pages}{675} (\bibinfo{year}{2018}), \eprint{1803.01853}.

\bibitem[{\citenamefont{Greljo and Marzocca}(2017)}]{Greljo:2017vvb}
\bibinfo{author}{\bibfnamefont{A.}~\bibnamefont{Greljo}} \bibnamefont{and}
  \bibinfo{author}{\bibfnamefont{D.}~\bibnamefont{Marzocca}},
  \bibinfo{journal}{Eur. Phys. J.} \textbf{\bibinfo{volume}{C77}},
  \bibinfo{pages}{548} (\bibinfo{year}{2017}), \eprint{1704.09015}.

\bibitem[{\citenamefont{Mangano et~al.}(2017)}]{Mangano:2016jyj}
\bibinfo{author}{\bibfnamefont{M.~L.} \bibnamefont{Mangano}}
  \bibnamefont{et~al.}, \bibinfo{journal}{CERN Yellow Report} pp.
  \bibinfo{pages}{1--254} (\bibinfo{year}{2017}), \eprint{1607.01831}.

\bibitem[{\citenamefont{Khachatryan et~al.}(2015)}]{CMS:2014jea}
\bibinfo{author}{\bibfnamefont{V.}~\bibnamefont{Khachatryan}}
  \bibnamefont{et~al.} (\bibinfo{collaboration}{CMS}), \bibinfo{journal}{Eur.
  Phys. J.} \textbf{\bibinfo{volume}{C75}}, \bibinfo{pages}{147}
  (\bibinfo{year}{2015}), \eprint{1412.1115}.

\bibitem[{CMS(2017)}]{CMS-PAS-SMP-16-019}
\bibinfo{type}{Tech. Rep.} \bibinfo{number}{CMS-PAS-SMP-16-019},
  \bibinfo{institution}{CERN}, \bibinfo{address}{Geneva}
  (\bibinfo{year}{2017}), \urlprefix\url{https://cds.cern.ch/record/2264556}.

\end{thebibliography}
\end{document}